\documentstyle[preprint,aps]{revtex}
\draft
\tighten
\begin{document}
\title{Universal Prefactor of Activated Conductivity in the Quantum Hall
Effect}
\author{D.\ G.\ Polyakov$^{(a)}$ and B.\ I.\ Shklovskii}
\address{Theoretical Physics Institute, University of Minnesota,
Minneapolis, Minnesota 55455}
\maketitle
\begin{abstract}
The prefactor of the activated dissipative conductivity in a plateau range
of the quantum Hall
effect is studied in the case of a long-range random potential. It is
shown that
due to long time it takes for an electron to drift along the perimeter
of a large percolation cluster, phonons are able
to maintain quasi-equilibrium inside the cluster. The saddle points
separating such clusters may then be viewed as ballistic point
contacts between electron reservoirs with different electrochemical
potentials.
The network of ballistic conductances is shown to determine
the conductivity.
The prefactor is universal and equal to 2$e^2/h$ at an integer filling
factor $\nu$ and to  2$e^2/q^{2}h$  at $\nu=p/q$.
\end{abstract}
\pacs{PACS numbers: 73.40.Hm}
\narrowtext

The temperature dependence of the activated dissipative
conductivity is widely used to study energy gaps in the quantum Hall
effect. When the Fermi level lies in the middle between two Landau
levels $\sigma_{xx}$ has a
form\cite{stahl85,clark88,clark90,sachrajda90,katayama,nicholas}
\begin{equation}
\sigma_{xx}=\sigma_0 \exp(-\Delta/T),
\label{sigma}
\end{equation}
provided the temperature $T$ is not too low (we use energy units for $T$).
The prefactor $\sigma_0$
has attracted a great deal of interest since it was
claimed \cite{clark88,clark90} that $\sigma_0$ is equal
to $e^2/h$ and does not depend on the Landau level number. This
puzzling universality was reported to be valid in the
fractional quantum Hall effect (FQHE) as well: at filling $\nu=p/q$ the
prefactor
$\sigma_0=(e/q)^2/h$ \cite{clark88,sachrajda90}. On the other hand,
both
the numerical value of the prefactor \cite{nicholas} and the very
independence of $\sigma_0$ on $T$
\cite{katayama} have been questioned.

There have been several interesting attempts to calculate the prefactor
universally, for any type of disorder, using the idea that at $T=0$
the extended state exists at only one energy and
phase-breaking processes are responsible for the delocalization of
electrons within a narrow band of states near the Landau level center
\cite{clark90,katayama,lee92,dassarma93,lee93,bhatt93}.
Nevertheless, it has not been shown yet that the range of
temperatures exists where $\sigma_0=e^2/h$.

We suggest a more specific approach and consider separately the
cases of a short and a long range random potential.
In our previous work \cite{polyakov94} we evaluated explicitly
two major
contributions to the conductivity in the case of a white noise
potential: the contribution of the narrow band of
delocalized
states which appear near the Landau level
center at finite temperatures and
the other one, related to the variable range hopping (VRH) between
localized states in the tail of the density of states.
It was shown that, in agreement with the conjecture made in Ref.
\cite{bhatt93}, the interplay of these
contributions is responsible for the inflection point in the
dependence of $\ln\sigma_{xx}$ on $1/T$ (Fig. 1a). We demonstrated
 that approximation of $\sigma_{xx}$ by Eq.~(\ref{sigma})
with $T$-independent $\sigma_0$ in the vicinity of the inflection
point yields
$\sigma_0\alt e^2/h$ being a very
slow function of the rate of phase breaking processes. So we concluded
 that, strictly speaking, no universality can be
obtained for the case of a short range potential.

In this paper, we calculate the prefactor $\sigma_0$ in the case
of a long range random potential. Such a potential exists in high
mobility
heterojunctions where the two-dimensional electron gas (2DEG) is
separated from randomly situated donors
by an undoped layer of width $d$ which is much larger than the
magnetic length
$\lambda$. The potential harmonics with wavelengths smaller than $d$
do not reach the 2DEG.
Therefore, the characteristic length of the potential fluctuations
is equal to $d$.
Most of experimental results for the prefactor $\sigma_0$
were obtained
on the structures with large $d$. Thus the consideration of a long
range potential is of special interest. We start from the integer quantum
Hall effect but generalization to the FQHE is straightforward.

We show below that in the case of a smooth potential due to the
suppression of tunneling, the conductivity may be described by a
completely {\em classical} theory which yields
a range of temperatures growing as a power of the ratio
$d/\lambda$, where the prefactor is constant and universal (Fig. 1b).
We obtain $\sigma_0=e^2/h$ for the contribution of one Landau level
to the conductivity in this range. The Plank constant appears in this
 expression only through the density of states of the Landau level.

We start with a demonstration of the universality of $\sigma_0$ for a
simple model problem. Instead of a random potential we consider
a periodic ``chess-board" one
\begin{equation}
W(x,y)=W \cos(\pi x/d) \cos(\pi y/d).
\label{w}
\end{equation}
It is assumed that the Fermi level $\mu$ is located in the middle
between two Landau levels separated by the gap $2 \Delta$
and that $W\alt \Delta$ and $d\gg \lambda$. We consider
the contribution to the conductivity
of electrons activated to the upper Landau level where the potential
energy oscillates around zero according to Eq.~(\ref{w}). At
$T\ll \Delta$ the concentration of activated electrons is exponetially
small so that we can neglect the interaction between them. The
inequality $d\gg \lambda$
allows us to treat the electrons semiclassically. Electron guiding
centers drift along equipotential
lines $W(x,y)=E$, all of which except one at $E=0$ are closed loops
(Fig. 2a). In principle, an electron can tunnel through a saddle
point separating two
loops of the same energy $E$. However, the probability of tunneling
falls off with increasing $|E|$ as $\exp(-|E|/T_{1})$,
where $T_{1}\sim W(\lambda/d)^{2}$\cite{fertig}.
Below we consider the wide temperature range $T_{1} \ll T\ll W$
in which tunneling can be neglected.

In the absence of inelastic processes $\sigma_0=0$. To find the
conductivity of electrons due to inelastic collisions, we should know
their energy distribution in the presence
of an external electric field.
Electrons with energies $E<0$ within a given well are in
equilibrium with each other because they circulate many times
around the well and experience many inellastic collisions before
leaving the well due to activation to trajectories with $E>0$.
Therefore, their distribution is
characterized by an electrochemical potential which is constant
inside the well. We show below that electrons with energies $0<E\alt
T$ play a very important role in transport. They occupy ``$T$-strips"
surrounding each potential hill (Fig. 2a). The distribution of these
electrons depends on the parameter $v\tau /d$, where $v$ is the
typical drift velocity and $\tau$ is the time necessary to change the
energy by $T$. Our crucial
assumption is that $v\tau\ll d$. This means that electrons
of $T$-strips ``tune" their electrochemical potential
so that it equals the electrochemical potential in the adjacent well.

Let us imagine that an electric field $F$ is applied along a diagonal
$MOQ$ of the chess-board (Fig. 2a). Then the
electrochemical potential drops along the line $QOM$ only in the vicinity of
the saddle
point $O$. The value of this drop is $eU=\sqrt{2}eFd$, where $U$ is the voltage
drop between $M$ and $Q$.
The net current from the well $M$ to the well $Q$ appears as the
difference between the two opposite drift flows in $T$-strips near
the saddle point $O$. They are shown by in Fig. 2a. In the
crossection $AOD$
electrons going from $Q$ to $M$ and back have chemical
potentials $\mu_{Q}=-\Delta+eU/2$ and $\mu_{M}=-\Delta-eU/2$
respectively. When electrons arrive from one well to the other
and continue to drift along the $T$-strips,
their chemical potential very quickly relaxes to the new value.

Thus we have arrived at the concept of a network of reservoirs
(potential wells) connected via ballistic contacts (saddle points)
shown in Fig. 2b. Let us show that the conductance $G$ of each of
these contacts has the form
\begin{equation}
G={e^{2}\over h}\exp(-\Delta/T).
\label{g}
\end{equation}
If the $x$-axis is directed along $AOD$ with the reference point at
$O$, the current from $Q$ to $M$ can be calculated as
\begin{equation}
I_{QM}=-e\int_{0}^{\infty} dx n(x) v(x),
\label{current}
\end{equation}
where
$n(x)=(1/2\pi \lambda^{2})\exp(-\Delta/T+eU/2T-W(x)/T)$
is the two-dimensional concentration of electrons and
$v(x)=(c/eB)\partial W/\partial x$
is their drift velocity.
Calculation of the integral gives
\begin{equation}
I_{QM}={e\over h} T \exp(-\Delta/T+eU/2T).
\label{current1}
\end{equation}
Similarly, for the current in the opposite direction we get
$I_{MQ}=-(eT/h)\exp(-\Delta/T-eU/2T).$
Calculating the net current in the Ohmic conditions ($eU/T\ll 1$)
we arrive at Eq.~(\ref{g}).
It agrees with general expression $G=\nu e^{2}/ h$, where $\nu $ is the
filling factor at the saddle point\cite{chklovskii}. In our case
$\nu=\exp(-\Delta/T)$.

The conductivity of the whole resistor network shown in Fig. 2b is
\begin{equation}
\sigma_{xx}=G={e^{2}\over h}\exp(-\Delta/T).
\label{sigmanet}
\end{equation}
Note that this calculation gives contribution to the conductivity
only of electrons on the upper Landau level. Holes on the lower
Landau level form their own resistor network.
It can be considered independent of the electron network so long as
the electron-hole recombination is, as usual, slow enough. Therefore,
the total conductivity of both parallel networks is of the form of
Eq.~(\ref{sigma}) with $\sigma_0=2e^2/h$.

Let us now turn to the random long range potential. The simplest
way to imagine it, departing from the chess-board potential, is to
assume that heights of the saddle points $W_{i}$ are randomly
distributed in an interval of energies $(-W_{1}, W_{1})$, where
$W_{1}$ is comparable with $W$. Assuming then that $v\tau\ll d$
and repeating the same arguments as for the chess-board case, we
arrive at the network of random conductances
\begin{equation}
G_{i}={e^{2}\over h}\exp(-\Delta/T-W_{i}/T).
\label{grandom}
\end{equation}
To calculate the conductivity of such a network, we use the
Dykhne
theorem \cite{dykhne}. According to
this theorem a two-dimensional  network of random conductances
with a symmetrical distribution of
 $\ln G_{i}$ around the average value $<\ln G_{i}>$ has the
conductivity
\begin{equation}
\sigma_{xx}=\exp(<\ln G_{i}>).
\label{sigmadykhne}
\end{equation}
In our case $<\ln G_{i}>=-\Delta/T$ because $<W_{i}>=0$, and thus
we again arrive at Eq.~(\ref{sigmanet}).
It is important that the limits of applicability of
Eq.~(\ref{sigmanet}) are much broader for a random potential than for the
periodic one with
the same values of $W$ and $d$. This happens because, according to
percolation theory, randomness of the saddle point heights
generates two new large scales at $|E|\ll W$:
\begin{equation}
\xi=d\left( {W\over |E|}\right)^{\nu_{p}},~~~~~p=d\left( {W\over
|E|}\right)^{\gamma}.
\label{xi-p}
\end{equation}
Here $\xi$ and $p$ are the ``diameter" and the perimeter of critical
equipotential loops at a given energy $E$. By ``critical" we mean
that the
probability of finding a loop with a size smaller than the critical one
at a given $E$ decays as a power law function of the size, while it
decays exponentially at sizes larger than critical. It is known that
$\nu_{p}=4/3$ and $\gamma=\nu_{p}+1=7/3$ \cite{isichenko}.
In the above derivation of Eq.~(\ref{sigmanet}) based on
Eq.~(\ref{sigmadykhne}), we considered all of the saddle points of the random
chess-board potential. Actually, the conductivity is determined only
by
those of them for which $|W_{i}| \sim T$. At $T\ll W$ these saddle
points separate critical loops corresponding to $|E|\sim T$ which have
very long perimeters
\begin{equation}
p_{T}\sim d\left( {W\over T}\right)^{\gamma}.
\label{pT}
\end{equation}
Such ``$T$-loops" together with saddle points with heights
$|W_{i}|\sim T$ form a network which is topologically similar to the
chess-board. We can consider it as a
network of ballistic conductances given by Eq.~(\ref{grandom}) and
arrive at Eq.~(\ref{sigmanet}) if the electrons of a $T$-loop are in
equilibrium with the cluster of potential wells they circulate
alongside, i.e. if $v\tau\ll p_{T}$. According
to Eq.~(\ref{pT}) this inequality puts a much weaker condition on
$\tau $ than $v\tau\ll d$, which is required for the periodic chess-board.

Let us discuss now the width of the interval of $T$ where $v\tau\ll
p_{T}$ and $\sigma_0=2e^2/h$. This requires an explicit discussion of
the nature of inelastic processes. As we mentioned above, activated
electrons are far from each other and thus interaction between them
is negligible. We can neglect also the interaction with electrons on
 the Fermi level: they are concentrated in droplets only in the rare
 places where the Fermi level touches the Landau level.
Therefore, we assume
below that inelastic processes are only due to electron-phonon
interactions. (Note that any additional process is able only to expand
the range of validity of the universal prefactor.)

To discuss the electron-phonon scattering,
we first estimate the typical drift velocity:
$v\sim cW/eBd\sim W\lambda^{2}/\hbar d$.
One can easily verify that under realistic conditions $v>s$, where $s$
is the sound velocity. This means that one-phonon processes are
permitted by the conservation laws. To evaluate the corresponding
electron-phonon collision time $\tau_{c}$, one may assume that an
electron is in a uniform electric field $\sim W/ed$. Then using Fermi's
golden rule, one obtains $\tau_{c}\sim \hbar/\alpha T$ for
$T\agt W\lambda/d$ and $\tau_{c}\sim\hbar d/\alpha \lambda W$ for
$T\alt W\lambda/d$.
Here $\alpha=\hbar C^2/2\rho
s^3\lambda^2$ is the electron-phonon coupling constant, $C$ and $\rho$
are the deformation potential
constant and the crystal density respectively. For $GaAs$ $\alpha\simeq
0.1(100\AA/\lambda)^2$ \cite{gantmakher}.
The characteristic energy $W\lambda/d$ appears above because
typical hops
occur between two trajectories a distance $\lambda$ from
each other and $W\lambda/d$ is the typical phonon energy. In order
to get $\tau$
from $\tau_{c}$ one should again consider separately the two cases
when
$W\lambda/d$ is larger or smaller than $T$. While in the former
case the energy changes by $T$ via one scattering, in the latter one
an electron slowly diffuses along the energy axis. This yields
\begin{equation}
\tau \sim {\hbar\over {\alpha T}}\left( {Td\over W\lambda}\right)^{2}
\label{tau1}
\end{equation}
for $T\agt W\lambda/d$ and
$\tau\sim \hbar d/ \alpha \lambda W$
for $T\alt W\lambda/d$. Using these results together with
Eq.~(\ref{pT}),
we find that for large enough spacer $d\gg d_{c}$, the inequality
$v\tau\ll p_{T}$ is valid for $T$-loops and, correspondingly,
$\sigma_0=2e^2/h$ in a wide
range of temperatures $T_{1}\ll T\ll T_{2}$.
Here $d_{c}=\lambda/\alpha^{1/(\gamma+1)}$ and
$T_{2}=W\alpha^{1/(\gamma+1)}$. For $\lambda\simeq$10 nm and
$\alpha\sim0.1$ one gets $d_{c}\simeq$20 nm. Thus for samples
with $d>>20$ nm we arrive at the universal prefactor
$\sigma_0=2e^2/h$. Below we deal only with the case $d\gg d_{c}$.
We will consider elsewhere the narrow range $\lambda<d<d_{c}$,
where the crossover between the short and long range cases takes
place.

Let us briefly discuss what happens away from the temperature
range $T_{1}\ll T\ll T_{2}$. If $T\alt T_{1}$ tunneling becomes
important and an additional VRH contribution to the conductivity
appears,
similar to the case of a short range potential \cite{polyakov94}. VRH
conductivity is determined by hops at a so called transport energy
which is a result of an interplay between the probabilities
of tunneling and of activation \cite{shapiro85}. The transport energy
is negative and its absolute value grows with decreasing $T$. Thus
the plot of $\ln\sigma(1/T)$ eventually deviates upwards  from the
straight line of the universal regime.
Such deviations are seen in some of the experimental data
\cite{clark90,katayama} but they seem to occur at higher $T$ than
we predict.

On the other side, when $T\agt T_{2}$ the perimeter $p_{T}$ of the
$T$-loops
becomes smaller than $v\tau$. In other words the $T$-strip becomes
so wide that only the low energy part of it is still in equilibrium
with
the adjacent potential well. In order to find the width $\Gamma$ of
the band $0<E\alt \Gamma$, where equilibrium is still supported,
one
should solve the equation
\begin{equation}
v\tau(\Gamma)\sim p(\Gamma).
\label{Gamma}
\end{equation}
 Here $\tau(\Gamma)$ is the time it takes to change
the energy by $\Gamma$ via diffusion in the energy space (for $T\gg
T_{2}$ and
$d\gg d_{c}$ only the diffusion regime is relevant). This time can be
found by
replacing $T^{2}$ in the numerator of Eq.~(\ref{tau1}) with
$\Gamma^{2}$. We obtain the perimeter $p(\Gamma)$ from
Eq.~(\ref{xi-p}) by substituting $\Gamma$ for $|E|$. The solution of
Eq.~(\ref{Gamma}) yields $\Gamma\sim T_{2}^{\beta} T^{1-\beta}$,
where $\beta=(\gamma+1)/(\gamma+2)=10/13$. Repeating
the calculation of the ballistic conductance of a saddle point similar to
that using Eq.~(\ref{current}), we have to restrict the integration to
the strip in which
$0<W(x)\alt \Gamma$. (Electrons with $\Gamma\alt E\alt T$
circulating around
the hills $A$ and $D$ have the same the electrochemical
potentials and therefore do not contribute to the
ballistic current at the saddle point $O$.)
This leads to the replacement
$T\rightarrow \Gamma$ in the prefactor of Eq.~(\ref{current1}). As a
result, we get for the prefactor of conductivity at $T\gg T_{2}$
\begin{equation}
\sigma_0 \sim {e^2\over h}\left({T_{2}\over T}\right)^{\beta}.
\label{PreflargeT}
\end{equation}
Note that Eq.~(\ref{PreflargeT}) is equivalent to the result of
calculation of the effective diffusion coefficient for a classical
advective-diffusive motion \cite{isichenko}.

Eq.~(\ref{PreflargeT}) predicts
deviation downwards from the straight line in the plot of
$\ln\sigma $ vs $1/T$
when $T$ becomes larger than $T_{2}$ (Fig. 1b). Strong
deviations of this type were observed in Refs.
\cite{clark90,katayama}. In some cases the dependence
of $\ln\sigma $ on $1/T$ even saturates. We believe that such
behavior can
be explained using Eq.~(\ref{PreflargeT}), only if the screening of the
long
range potential by activated electrons and holes, which reduces $W$
and $T_{2}$, is taken into account. Screening becomes important
when the concentration of activated carriers
$(1/2\pi \lambda^{2})\exp(-\Delta/T)$ is comparable to the
fluctuations of the charge donor concentration which have length scale $d$. For
the case of large $d$, the latter concentration may be much smaller
than $1/(2\pi \lambda^{2})$, and therefore screening may become
important even at $T\ll \Delta$.

In conclusion, we have shown that in a heterostructure with a large
spacer there is a range of temperatures at which the prefactor
$\sigma_0$ is universal. It is equal to $2e^2/h$ if the Fermi
level is in the middle between two Landau levels.  In the case of the FQHE our
theory can be applied to the conductivity of excitations with fractional
charges
(quasielectrons and quasiholes). As a result
$\sigma_0=2e^2/q^2h$ at $\nu=p/q$. Our values of $\sigma_0$ differ by the
factor
 2 from the values
claimed in Refs. \cite{clark88,clark90}. Note, however, that larger
values of $\sigma_0$ were reported by another
group\cite{nicholas}. Moreover, recently $\sigma_0$ has been  found
to be proportional to $1/T$ \cite{katayama}. We do not know how to
resolve these contradictions. It seems that more experimental evidence is
needed here.

We are grateful to I.\ L.\ Aleiner, M.\ M.\ Fogler and V.\ I.\ Perel for
valuable discussions.
This work was supported by the NSF Grant No.\ DMR-9321417.

\protect\begin{figure}
\caption{ Schematical plot of $\ln\sigma $ vs $1/T$ when the Fermi level
is in the middle between two Landau levels: (a) as calculated
for a short range potential\protect\cite{polyakov94}, (b) as obtained in this
work for a long range potential.}
\label{Fig. 1}
\end{figure}

\protect\begin{figure}
\caption{ (a) Equipotential lines $W(x,y)=0$ (square lattice) and
$W(x,y)=T$ for a chess-board potential. Saddle point $O$ separates
two hills $A$, $D$ and two wells $Q$, $M$. Arrows show directions of
the drift current.
(b) Equivalent curcuit. Each conductance $G$ is given by
Eq.~(\protect\ref{g}).}
\label{Fig. 2}
\end{figure}

\end{document}